# Meta-Entanglement

**Paola Zizzi**
*Department of Brain and Behavioural Sciences, University of Pavia,*
*Piazza Botta, 11, 27100 Pavia, Italy*

**Abstract**

We give a meta-logical interpretation of the entanglement mechanism of quantum space-time in terms of the sequent calculus of a quantum sub-structural logic. This meta-logical picture is based mainly on the two meta-rules cut and EPR, and on the new meta-theorem "teleportation" (TEL), built by the use of the above meta-rules, both performed in parallel. The proof of (TEL)-theorem fairly reproduces the protocol of quantum teleportation. In the framework of space-time entanglement, the conclusion of the (TEL)-theorem is that the entangled space-time can convey the quantum teleportation of an unknown quantum state.
We also introduce two new structural rules: the Hadamard (H)-rule and the CNOT-rule, the latter being used, together with the cut, in the proof of the new theorem "Entanglement" (ENT).

**1. Introduction**

The goal of this paper is to give a formal meta-logical description of the entanglement mechanism of quantum space-time [1]. In fact, the entangled quantum space-time looks like a quantum control (a quantum meta-language [2]) over a quantum object-language, the latter being used by a quantum computer.

As it is well known, quantum entanglement is one of the most important features of quantum computing [3] as it leads to massive quantum parallelism, hence to exponential computational speed-up. In a sense, quantum entanglement is considered as an implicit property of quantum computation.

In [1] we described the entanglement mechanism of quantum space-time itself. The study of such a mechanism was based on various concepts and results already illustrated in previous works, among which the discrete, quantum version of the empty space-time of the de Sitter universe [4], its logical quantum-computational realization in terms of a Quantum Growing Network (QGN) [5] and the quantum extension [6] of the holographic principle [7] [8].

In [1] we made a change to the QGN, by including an observer who, standing on the nth. horizon of the de Sitter's discrete universe, observes the ($n$-1) th. horizon by means of a photon with the appropriate energy. The presence of the observer is equivalent to add a projector to node $n$, where there was already a Hadamard quantum gate. The apparent loss of the quantum information due to the measurement is restored by the quantum gate CNOT also added to node n, which entangles a qubit of node n with one of node $n$-1, by using the bit obtained from the measurement as target. This new quantum network was called OQGN, where "$O$" stands for "Observer".

At the light of such results, quantum space-time then appears to behave like a quantum computer, although at a different logical level. While a physical quantum computer (QC) uses a quantum logical language, entangled space-time uses a quantum meta-language (QML), which controls the QC logical language as a quantum object-language (QOL). The QML cannot be given to the quantum physical machine.

The entangled quantum space-time of our model is empty, therefore it cannot support physical quantum logic gates like the CNOT etc. Entangled space-time is endowed with QML, which can be reflected into the QOL of the quantum network. This is realized by means of two meta-rules, a new theorem, the "Entanglement" (ENT)-theorem, a new meta-theorem, the "Teleportation" (TEL)-theorem, and two new structural rules, the Hadamard (H)-rule and the CNOT-rule.

The meta-rules are rules that describe how other rules should be used, and thus belong to the realm of meta-logic. The meta-rules cannot be given to a machine (not even to a quantum computer). So far, only two meta-rules have been known in sequent calculus: the cut rule [9] and the EPR rule [2]. The latter, which has been discovered quite recently [2], was built by the use of the quantum logical connective "Entanglement" in a quantum version [2] of Basic Logic [10].

Weak (sub-structural) logics are those logics that do not have at least one of the (usual) structural rules: Weakening (W), Contraction (C), and Permutation (P). These logics, like linear logic [11] and Basic logic, are most suitable for computer science. In fact, in logic, a weak structure leaves more room for new connectives (like the two new quantum connectives [2] "quantum superposition" and "entanglement"), for meta-rules, and, as we will see, also for new structural rules, which are best suited for quantum computing.

While the logical rules introduce a new logical formula either on the left or on the right of the sequent, the structural rules operate on the structure of the sequent itself. Basic logic, the logic upon which our quantum-computational logic was built [2], is sub-structural as it has not the structural rules C and W. The absence in Basic logic of the C and W rules corresponds to the validity of the no-cloning [12] and no-erase [13] theorems, respectively, in quantum computing. This correspondence was found in [2]. The above no-go theorems just state that there is not a unitary operation (performed by a quantum gate), which can reproduce quantum copying (and erase). Then, the sequent calculus for a sub-structural logic is in agreement with unitary operations in quantum computing. Extrapolating a little further, we argue that the structural rules allowed in a quantum-computational logic, are those that describe quantum logic gates. In a sense, the quantum version [2] of Basic logic is not sub-structural *tout curt*, but it is sub-structural only with respect to two usual structural rules, which are not in agreement with quantum computing. In this paper, we will introduce two new structural rules, which describe the action of two important quantum logic gates, the one qubit gate Hadamard (H), which creates superposition, and the two qubits gate CNOT, which creates entanglement.

The cut rule can be used in sequent calculus to describe the projective measurement of a qubit [2].

In the same way, the EPR rule can be used to describe the projective measurement of a Bell state [2] [14].

In this sense, the two rules cut and EPR are meta-rules, and for this reason they cannot be given to a quantum computer, which performs only unitary (reversible) operations.

The cut, the EPR, the H and the CNOT rules can be performed in "branches" (in parallel).

We consider the "branched" cut rule, which corresponds to a reversible quantum measurement [15] of a qubit, as performed by an observer who uses an internal logic [16]. In this case, the "internal" observer appears to be in a non commutative space, which is a one-to-one correspondence with the one qubit state [17], namely, the fuzzy sphere [18].

In the same way, the "branched" EPR rule, corresponds to a reversible measurement of a Bell state [19], as performed by an internal observer. The latter is in a quantum (non commutative) space, which is in a one-to-one correspondence with the Bell state. The quantum simulation of entangled space-time on a fuzzy sphere is under study [20].

The structural CNOT-rule describes, in sequent calculus, the unitary (reversible) operation performed by a CNOT gate, which creates entanglement. The CNOT gate uses one classical bit (for example 0) as target, and a cat state as control. When the CNOT rule is performed in parallel, it corresponds to the simultaneous use of bits 0 and 1 as targets. The result is that there is no entanglement in the conclusion of the proof.
The CNOT-rule turns to be fundamental in the proof of the new theorem "Entanglement"(ENT). The proof of the (ENT)-theorem reproduces all the steps needed in the entanglement mechanism of two pixels of Planck area belonging to two subsequent spatial slices at two successive Planck time steps, as described in [1].
Also, we find that the (ENT)-theorem, when demonstrated in parallel, does not give entanglement in the conclusion. The consequence, in the case of the entangled space-time, is that two pixels of Planck area cannot get entangled to each other if they belong to the same spatial slice, at a given Planck time step. This extra result can be seen as a no-go theorem, in accordance with the no self-entanglement theorem [1] [2] when interpreted in the case of space-time entanglement.
Finally, we state the meta-theorem "Teleportation" (TEL), and give a formal proof of it in terms of the two (branched) meta-rules cut and EPR. A meta-theorem is proven within a meta-theory, and may reference concepts that are present in the meta-theory but not in the object-theory.
The theorem TEL is a meta-theorem because it deals only with two meta-rules (the cut and the EPR) rather than with logical or structural rules, i.e., it is a theorem within the meta-theory. We show that the proof of the (TEL)-meta-theorem fairly reproduces the protocol of quantum teleportation. Moreover, in the case of space-time entanglement, such a theorem suggests that two maximally entangled pixels of Planck area can convey the teleportation of the unknown quantum state of a particle. At the end, the unknown quantum state is entangled with a pixel of Planck area.
The fact that the proof of the (TEL)-meta-theorem can only be performed in parallel, means that it corresponds to the operation identity. In fact we start from entanglement in the premise, and we end with entanglement in the conclusion, although one party of the original EPR pair was replaced with the unknown quantum state.
In this paper, for the sake of brevity, we will not give a full review of the vast underlying theoretical background. In case, we will summarize only those topics that could be particularly useful for an easier approach to new results.
The paper is organized as follows.
In Sect. 2, we will review the cut rule (and its interpretation as a projective measurement), the connective "entanglement", and the EPR rule. Also, we will describe the cut rule in parallel as a reversible quantum measurement.
In Sect. 3, we will introduce the branched EPR rule, and the two new structural rules, the Hadamard (H)-rule and the CNOT-rule, together with their branched versions.
In Sect. 4, we will state the Entanglement (ENT)-theorem, and the Teleportation (TEL)-meta-theorem.
Sect. 5 is devoted to the conclusions.
In the Appendix, the interested reader can find the logical derivations and the proofs of the theorems in the formalism of sequent calculus.

## 2. A short review of some basic topics and notations

In this Section, we will review some basic topics, among which the cut rule and its interpretation as the projective measurement of a superposed quantum state. Moreover, we will illustrate the "branched" cut rule, which does not destroy quantum superposition. Also, we will review the quantum connective "entanglement", and finally, the EPR rule.
To start, let us recall some notations of sequent calculus (for more details see, for example, Ref. [2] [10]).
The symbol $\vdash$ known as "turnstile", separates the assumptions on the left from the propositions on the right. A and B denote propositions, and $\Gamma, \Delta$ are finite (possibly empty) sequences of formulae, called contexts.
We say that the finite list $\Delta$ of assertions follows from a finite list of assertions $\Gamma$ (or equivalently "$\Gamma$ yields $\Delta$") and write: $\Gamma \vdash \Delta$, where $\vdash$ ("yields" ) is a metal-link between assertions. $\Gamma$ is said to be the antecedent, and $\Delta$ the consequent of the sequent $\Gamma \vdash \Delta$. Either $\Gamma$ or $\Delta$ (or both) can be empty. If the consequent $\Delta$ is empty: $\Delta \equiv \emptyset$ this is interpreted as false, that is $\Gamma \vdash$ means that $\Gamma$ proves falsehood, and therefore it is inconsistent. Instead, an empty antecedent $\Gamma \equiv \emptyset$ is assumed true, that is, $\vdash \Delta$ means that $\Delta$ follows without any assumption, that is, it is always true. We say that $\vdash \Delta$ is a logical assertion.
In particular, the sequent $\Gamma \vdash A$ means " the sequence $\Gamma$ yields $A$". The sequent $\vdash A$ is the assertion "$A$ is true". The primitive negation of $A$ is indicated by $A^{\perp}$, and the sequent $\vdash A \,\&\, A^{\perp}$, which means "$A$ and $A^{\perp}$ are both true", is interpreted as the cat state $|Q\rangle_A = \frac{1}{\sqrt{2}}\left(|0\rangle_A + |1\rangle_A\right)$ (where the propositions $A^{\perp}$ and $A$ are interpreted as the bits $|0\rangle_A$ and $|1\rangle_A$, respectively). The logical interpretation of the qubit $|\!-A \,\&\, A^{\perp}$ evidently invalidates the non contradiction principle $A \,\&\, A^{\perp}|\!-$ . The latter is in fact invalidated in paraconsistent logic [21] and in Basic logic [10].

**2.1 The cut rule**

The rule of cut is a meta-rule, which cannot be given to a machine, as, for example, a computer. The cut rule states that, when a formula A can be concluded and this formula may also serve as a premise for concluding other statements, then the formula A can be "cut out" and the respective derivations are joined. When a proof bottom-up is constructed, this creates the problem of guessing A (since it does not appear at all below). This issue is addressed in the theorem of cut-elimination [9]. The cut rule is:

$$\frac{\Gamma|\!-A \qquad \Delta, A|\!-B}{\Gamma, \Delta|\!-B} \quad . \tag{2.1}$$

In Basic logic [10] it holds logical "visibility" of formula (or absence of active contexts). One defines active contexts those which are close to the formula and passive context those which are separated from the formula by the sequent. Then, in basic logic the cut rule reads:

$$\frac{\Gamma|\!-A \qquad A|\!-B}{\Gamma|\!-B} \tag{2.2}$$

Because of logical visibility, it follows that in basic logic, by no means a superposed state can be entangled with the environment, unless a projective measurement is performed.

In our case, one might figure out passive contexts as those which belong to the computational state while active contexts will be those which do not, like external observers, measurement apparatus, the environment, in summary the external classical world.

The cut rule (2.2) in Basic logic destroys the superposed state $|\!- A \,\&\, A^{\perp}$ . This will be shown in the Appendix.

The cut rule in parallel (or "branched" cut rule) instead preserves superposition. This also will be shown in the Appendix. .

The cut rule corresponds to a projective measurement, not to a unitary operator. When performed in parallel, the cut rule corresponds to the superposition of two projective measurements (the Cat-mirror measurement $M_C$, a special case of the mirror measurement [15]) and the original qubit $|\Psi\rangle$ is left unchanged.

The Cat-mirror measurement is given below.

$$\mathrm{M_C}|\Psi\rangle = |\Psi\rangle$$

where

$$\mathrm{M_C} = \left(M_0 + M_1\right) = I_2$$

is the Cat-mirror operator,

$$|\Psi\rangle = \frac{1}{\sqrt{2}}\left(|0\rangle + |1\rangle\right)$$

is the one qubit cat state, and $\mathrm{M}_0$, $\mathrm{M}_1$ are the two orthogonal projectors of $C^2$, and $\mathrm{I}_2$ is the identity in $C^2$:

$$\mathrm{M}_0 = |0\rangle\langle 0| \,, \quad \mathrm{M}_1 = |1\rangle\langle 1| , \quad I_2 = \begin{pmatrix} 1 & 0 \\ 0 & 1 \end{pmatrix} \quad .$$

This reversible measurement can be interpreted as performed by a (fictitious) internal observer in a non-commutative space, which is in a one-to-one correspondence with the qubit [16] [17].

**The logical connective @ = "entanglement"**

Entanglement is a strong quantum correlation, which has no classical analogous. Then, the logic having room for the connective "entanglement", should be selected as the most adequate logic for quantum mechanics, and, in particular, for quantum computing. Quantum entanglement is mathematically expressed by a particular superposition of tensor products of basis states of two (or more) Hilbert spaces such that the resulting state is non-separable. For this reason, the new logical connective, which describes entanglement, is both additive and multiplicative.
We introduced the connective @ = "entanglement" [2] by solving its definitional equation, and we got the logical rules for @. It turned out that @ is a (right) connective given in terms of the (right) additive conjunction & and of the (right) multiplicative disjunction $\wp = \text{"}par\text{"}$.
Bell states will be expressed, in logical terms, by the expression $Q_A @ Q_B$.
Then, the logical structure for, say, the Bell state $\left|\Phi_+\right\rangle_{AB}$ is:

$$\vdash (A \wp B) \& (A^\perp \wp B^\perp)$$

where we recall that the connective $\wp$ = "par" is the multiplicative conjunction on the right in Basic logic and linear logic.
Similarly, the logical structure for the Bell state $\left|\Psi_+\right\rangle_{AB}$ is:

$$\vdash (A \wp B^\perp) \& (A^\perp \wp B)$$

In the following, we will consider only the logical expression for $\left|\Phi_+\right\rangle_{AB}$, as the case for $\left|\Psi_+\right\rangle_{AB}$ is obtained by exchanging $A$ with $A^\perp$.
Eventually, we get the following definition:
The two composite propositions $Q_A \equiv A \,\&\, A^\perp$ and $Q_B \equiv B \,\&\, B^\perp$, will be said maximally entangled if they are related to each other by the logical connective @.
The definitional equation of the logical connective @ = "entanglement" will be given in the Appendix, together with its rules and properties.

### 2.5 The EPR-meta-rule

As it is well known, if two quantum systems $S_A$ and $S_B$ are entangled, they share a unique quantum state, and even if they are far apart, a measurement performed on $S_A$ influences any subsequent measurement performed on $S_B$ (the EPR paradox).
Let us consider Alice, who is an observer for system $S_A$, which is the qubit $Q_A$, that is, she can perform a measurement of $Q_A$. There are two possible outcomes, with equal probability 1/2:
i) Alice measures 1, and the Bell state collapses to $\left|1\right\rangle_A \left|1\right\rangle_B$ .
ii) Alice measures 0, and the Bell state collapses to $\left|0\right\rangle_A \left|0\right\rangle_B$.
Now, let us suppose Bob is an observer for system $S_B$ (the qubit $Q_B$). If Alice has measured 1, any subsequent measurement of $Q_B$ performed by Bob always returns 1. If Alice measured 0, instead, any subsequent measurement of $Q_B$ performed by Bob always returns 0.
To discuss the EPR paradox in logical terms, we introduced the EPR rule [2], whose derivation is given in the Appendix.

## 3. The EPR-rule in parallel, the H-rule and the CNOT-rule

In this Section, we will introduce the "branched" EPR-rule (EPR-rule in parallel) describing the action of a fictitious observer living in a quantum space, which is in a one-to-one correspondence with the Bell state. Also, we introduce two new structural rules, the Hadamard (H)-rule, and the CNOT-rule, which describe quantum superposition and entanglement, respectively. Also, we will show that the "branched" versions of the (H) and CNOT rules do not give quantum superposition and entanglement, respectively, in the conclusion.

### 3.1 The "branched" EPR-rule

In this section, we will introduce the "branched" EPR-rule. We will show that two EPR applied in parallel give back entanglement.
The two EPR in parallel follow from two cuts in parallel, an internal measurement, described by a unitary operator. Then, the cut in parallel is a reversible operation, and gives back the original entangled state. The proof is given in the Appendix.
Once the cut is not performed in parallel, it corresponds to an external, irreversible operation. This suggests that the observer performing a quantum measurement not in parallel of qubit $Q_A$ at node $n = 0$, at time $t_0 = t_P$ (where $t_P$ is the Planck time) must be considered as external to node $n = 0$, in fact she stands on node $n = 1$ at time $t_1 = 2t_p$ [1]. In this framework, then, an external observer results as delayed in time.
The action of the EPR rule is similar to that of the cut, but while the cut is interpreted as a projective measurement on a one qubit state, the EPR is interpreted as a Bell measurement.
The EPR rule, when applied in parallel, gives back the Bell state. It corresponds to a "Bell mirror measurement" (a special case of Mirror measurement for Bell states [19]). Let us define the "Bell mirror measurement":

$$\mathrm{M_B}|\Psi\rangle = |\Psi\rangle,$$

where

$$|\Psi\rangle = \frac{1}{\sqrt{2}}(|00\rangle + |11\rangle)$$

is the Bell state,

$$\mathrm{M_B} = (M_0 + M_1) \otimes I_2 = I_2 \otimes I_2 = I_4$$

is the Bell mirror operator, $I_2$ is the identity in $\mathrm{C}^2$, and $I_4$ is the identity in $\mathrm{C}^2 \otimes C^2 = C^4$.

### 3.2 The Hadamard (H)-rule

The quantum gate Hadamard (*H*) is a one-qubit logic gate. It transforms a classical bit into a cat state. When *H* is applied to the bit $|0\rangle$, it gives the (symmetric) cat state $|+\rangle = \frac{1}{\sqrt{2}}(|0\rangle + |1\rangle)$, instead when *H* is applied to the bit $|1\rangle$, it gives the (anti symmetric) cat state $|-\rangle = \frac{1}{\sqrt{2}}(|0\rangle - |1\rangle)$. As *H* is unitary, it exists its inverse $H^{-1}$.
This means that quantum superposition is reversible:

$$H^{-1}|+\rangle = |0\rangle \qquad H^{-1}|-\rangle = |1\rangle.$$

When *H* is performed in parallel on bits $|0\rangle$ and $|1\rangle$, the resulting operation gives a bit $|0\rangle$:

$$\frac{1}{\sqrt{2}}(H|0\rangle + H|1\rangle) = |0\rangle.$$

In this section, we will introduce a new structural rule, the *H*-rule, which describes, in terms of sequents, the action of the Hadamard quantum gate. To this purpose, we would like to remind a few notions and definitions.
The classical bit $|0\rangle$ is interpreted as the sequent $\vdash A^{\perp}$ and the classical bit $|1\rangle$ as the sequent $\vdash A$. The two atomic assertions $\vdash A^{\perp}$ and $\vdash A$ are both asserted with certainty, and their truth values are both equal to 1.

However, bits $|0\rangle$ and $|1\rangle$ in the cat state $|+\rangle$ (or $|-\rangle$) are indicated by the sequents $\vdash^{\frac{1}{\sqrt{2}}} A^{\perp}$ and $\vdash^{\frac{1}{\sqrt{2}}} A$ (or $\vdash^{-\frac{1}{\sqrt{2}}} A$) respectively. The upper fixes $\pm\frac{1}{\sqrt{2}}$ on the sequents (the probability amplitudes in QM) are the "assertion degrees" [2] of the two assertions. Note that in this case we are considering cat states, then the two assertion degrees are real numbers, but of course in general they are complex numbers. The squared absolute values of the assertion degrees (probabilities in QM) are the partial truth values of the atomic assertions. The partial truth values of $\vdash^{\frac{1}{\sqrt{2}}} A^{\perp}$ and $\vdash^{\frac{1}{\sqrt{2}}} A$ are both $\frac{1}{2}$.

The symmetric cat state $|+\rangle$ corresponds to the compound assertion $\vdash A^{\perp}{}_{\frac{1}{\sqrt{2}}}\&_{\frac{1}{\sqrt{2}}}A$ (where ${}_{\alpha}\&_{\beta}$, with $\alpha,\beta\,\varepsilon\,C$, is a quantum logical connective named "superposition" [2]), while the anti-symmetric cat state $|-\rangle$ corresponds to the compound assertion $\vdash A^{\perp}{}_{\frac{1}{\sqrt{2}}}\&_{-\frac{1}{\sqrt{2}}}A$. The truth value of the compound proposition $\vdash A^{\perp}{}_{\alpha}\&_{\beta}A$ is 1.
Also, it should be noticed that it holds:

$$\vdash A{}_{\frac{1}{\sqrt{2}}}\&_{-\frac{1}{\sqrt{2}}}A\equiv\vdash\emptyset,$$

where Ø stands for the null proposition, and:

$$\vdash A\,\&A\equiv\vdash A.$$

The *H*-rule is that rule, which, starting from the single premise $\vdash A^{\perp}$ gives as consequence $\vdash A^{\perp}{}_{\frac{1}{\sqrt{2}}}\&_{\frac{1}{\sqrt{2}}}A$, or starting from the single premise $\vdash A$, gives as consequence $\vdash A^{\perp}{}_{\frac{1}{\sqrt{2}}}\&_{-\frac{1}{\sqrt{2}}}A$.
The logical derivation of the H-rule is given in the Appendix.
The *H*-rule is a structural rule, which is important for describing the original QGN [5] in terms of sequent calculus. In fact, we remind that at each node *n* of the QGN there is a *H* gate, which transforms a bit $|0\rangle$ (a connecting link) into a cat state $|+\rangle$ (an outgoing free link).

### 3.3 The CNOT-rule

The logical CNOT gate is a two-qubits quantum gate. It uses a classical bit (for example $|0\rangle_A$) as target, and a cat state $\frac{1}{\sqrt{2}}(|0\rangle+|1\rangle)_B$ as control. When the control is $|0\rangle_B$, the CNOT leaves the target unchanged, and when the control is $|1\rangle_B$, it flips the target. The result is a Bell state, for example: $|\Phi_+\rangle_{AB}=\frac{1}{\sqrt{2}}(|0\rangle_A|0\rangle_B+|1\rangle_A|1\rangle_B)$ .
(3.1)
The CNOT-rule is a structural rule, expressed by the sequents:

$$\frac{\vdash B,A}{\vdash B,A^{\perp}}CNOT\quad\text{(a)}\qquad\frac{\vdash B^{\perp},A}{\vdash B^{\perp},A}CNOT\quad\text{(b)}\ . \tag{3.1}$$

By exchanging $A\rightarrow A^{\perp}$, we have the equivalent formulation of the CNOT-rule:

$$\frac{\vdash B,A^{\perp}}{\vdash B,A}CNOT\quad\text{(a')}\qquad\frac{\vdash B^{\perp},A^{\perp}}{\vdash B^{\perp},A^{\perp}}CNOT\quad\text{(b')}\ . \tag{3.2}$$

The derivation of the C-NOT rule in Eqs. (3.1) and (3.2) is given in the Appendix.

To reproduce the whole action of the CNOT gate, which uses the qubit $|Q\rangle_B$ as control, and the bit $|0\rangle_A$ (or $|1\rangle_A$) as target, it should be considered, in the premises, the sequent: $\vdash Q_B, A^{\perp}$ (or $\vdash Q_B, A$), together with the &-refl : $Q_B \vdash B$ $Q_B \vdash B^{\perp}$. Entanglement is obtained from Eq. (3.1), or from Eq. (3.2), by using the formation rule of the connective "entanglement". For the derivation of the C-NOT rule action, which is given in the Appendix, we will consider Eq. (3.2).

### 3.4 The C-NOT rule in parallel

It is worthwhile considering the CNOT-rule in parallel, because, as we will see in the following, it allows to formulate a no-go theorem.

The CNOT-rule in parallel is given in the Appendix.

The conclusion of the C-NOT rule in parallel is $\vdash Q_B, Q_A$, which is the meta-logical description of $|Q\rangle_A \otimes |Q\rangle_B$. This is a separable state, therefore entanglement is lost when the CNOT-rule is performed in parallel.

## 4. Two new theorems, and a no-go theorem

In this Section, we will state two new theorems in sequent calculus, and will give the proof of each of them. The first is the "Entanglement" (ENT)-theorem, which describes the entanglement mechanism of quantum space-time. The second is the "Teleportation" (TEL)-meta-theorem, which describes how the entangled space-time conveys the teleportation of an unknown quantum state. Also, we will state and demonstrate a no-go theorem, which forbids entanglement in parallel.

### 4.1 The "Entanglement" (ENT)-theorem

The entanglement theorem states that from the premises:

$$\vdash Q_B, Q_A \qquad Q_A \vdash A^{\perp} \tag{4.1}$$

it follows:

$$\vdash Q_B @ Q_A \tag{4.2}$$

The proof of this theorem will be given in the Appendix.

However, it would be useful to clarify first what the premises and the conclusion of the (ENT)-theorem mean in physical terms. The first sequent $\vdash Q_B, Q_A$ in (4.1) means that two qubits $Q_A$ and $Q_B$ are given in a separable state, and the second sequent $Q_A \vdash A^{\perp}$ in (4.1) means that a projective measurement $P_0$ is performed on the qubit $Q_A$. The conclusion $\vdash Q_A @ Q_B$ in (4.2) means that then the two qubits $Q_A$ and $Q_B$ became entangled.

In [1] we described the entanglement mechanism of space-time in terms of a quantum growing network (QGN). In that scenario, nodes were a couple of quantum gates, the Hadamard gate H and the CNOT gate, plus a projector $P_i$ $(i = 0, 1)$. There were $2n+1$ links outgoing from each node n. Also, there were $2n+1$ connecting links. The connecting links did represent bits of classical information, and were transformed by H into qubits (outgoing links) at each node n. An observer standing on node n, did perform a projective measurement $P_i$ on one qubit $Q_A$ of node $n-1$, obtaining, for example the bit $|0\rangle_A$ for $i = 0$. Thereafter, the CNOT gate on node *n* did entangle a qubit $Q_B$ of node n with a qubit $Q_A$ of node $n-1$, by using $Q_B$ as control, and the measured bit $|0\rangle_A$ as target.

The meta-logical interpretation of such a mechanism of entanglement is the (ENT)-theorem, whose proof is given in the Appendix.

$\vdash Q_B @ Q_A$.

From the above results it is straight ford to derive a no-go theorem: No-Entanglement in parallel

In the Appendix, we will perform the cut in parallel in the proof of the (ENT)-theorem and we will show that we get back $\vdash Q_B, Q_A$, that is, a separable state in the conclusion.

To deeply understand the CNOT-rule in parallel, one should consider the non commutative space, which is in a one-to-one correspondence with a two qubits quantum state [20].

### 4.3 The "Teleportation" (TEL)-meta-theorem

In what follows, we will give a quantum meta-interpretation of the quantum teleportation protocol [23]. We call it the (TEL)-meta-theorem. It states that from the premises:

$\vdash (Q_A @ Q_B), Q_C \quad Q_A, Q_C \vdash^{\beta} C$ in parallel with $\vdash (Q_A @ Q_B), Q_C \quad Q_A, Q_C \vdash {}^{\alpha}C^{\perp}$

it follows:

$\vdash Q_C @ Q_B$

The (TEL)-theorem is a meta-theorem because, as we will see in the proof, it references only to the two meta-rules that are present in the meta-theory, namely the cut and the EPR rules, but not to logical or structural rules of the object-theory.
The proof will be given in the Appendix, but first let us "translate" the logical sequents in physical terms.
Alice (A) shares her qubit $Q_A$, with Bob (B), who owns the qubit $Q_B$, in a maximally entangled state (a Bell state) $Q_A @ Q_B$. Alice also has a qubit $Q_C$ in an unknown state. The total state is then $(Q_A @ Q_B), Q_C$. Then Alice performs a Bell measurement on $Q_A$ and $Q_C$ together. The Bell measurement is a joint quantum-mechanical measurement of two qubits that determines which of the four Bell states the two qubits are in. In sequent calculus this is achieved by performing two cuts in parallel, and then two EPR-meta rules in parallel. In the conclusion, the qubit $Q_A$ has disappeared, and the qubit $Q_C$ became entangled with the qubit $Q_B$, namely $Q_C @ Q_B$. This is illustrated in the proof of the "Teleportation" (TEL)-meta-theorem.
The quantum teleportation of the unknown qubit $Q_C$ can then be fairly described in terms of sequent calculus thanks to the quantum meta-language, which allows the introduction of the connective "quantum superposition". In the framework of the entanglement mechanism of space-time, the EPR-theorem says that the entangled space-time can convey the teleportation of the unknown quantum state of a particle. A pixel of Planck area encodes a qubit by the quantum holographic principle [6]. It has been shown [1] that two entangled Planck pixels belonging to two successive spatial slices form a Bell state. Let us call the two entangled pixels A and B, and suppose that a particle is in a unknown quantum state C, at the same site and time of pixel A. The unknown quantum state can be teleported from pixel A to pixel B and get entangled with B. From that we deduce that the unknown quantum state of a particle can get entangled with quantum space-time.
The peculiarity of the meta-teleportation theorem, is of being built directly in parallel (it could not be demonstrated otherwise, in sequent calculus). It looks like there is no a physical quantum space-time background to be taken into account in the process of quantum teleportation. However, it is just the entangled space-time (EST), as described in [1], which conveys the teleportation of the unknown quantum state C. This fact may lead to an apparent paradox: the EST is the agent for TEL, but quantum teleportation occurs without the need of an explicit EST-background. We will discuss that in more details in the next sub-sections.

### 4.4 Entanglement vs Quantum Teleportation

As far as entanglement is concerned, the EST cannot be put directly in a one-to-one correspondence with the EPR pair. In fact, the EST is a non connected space (a part from tiny wormholes connecting pairs of pixels of successive spatial slices). The QC state space of the EPR pair, instead, is isomorphic to a connected quantum space, namely, a fuzzy sphere. Therefore, by starting from the geometrical mechanism of entanglement of the physical space-time it is not possible to perform a one-to-one correspondence with the quantum system of the EPR pair. It follows that the quantum metalanguage (QML) endowed by the EST must be "reflected" onto the quantum object language (QOL) of the quantum system. This logical reflection then indicates a different topology of the two spaces. In summary, the entangled space-time (EST) is not on its own isomorphic to the quantum computer (QC) state space. If it was so, the quantum metalanguage of the EST would be the same as the logical quantum language of the QC. In that case, all theorems (comprising the ENT-theorem) would be meta-theorems like the TEL-theorem. This difference between the two languages is fortunate, because it allows entanglement in the quantum world. In fact, as we have seen, if we perform the CNOT in parallel, entanglement is lost. And performing the CNOT in parallel, means that we have built an isomorphism between the EST and the QC state space.
In the case of quantum teleportation, instead, the QC state space is disconnected because of the presence of the third qubit, $Q_C$. There is a natural isomorphism between the EST and the QC state space in the case of quantum teleportation, and there is no need of a logical "reflection". Note that in logic, the third qubit acts as a context. In principle, a context would not be admissible in Basic logic, as well as in its quantum version. However, $Q_C$ is a very peculiar kind of context, because it loses its role of context in the logical derivation, becoming part of the compound proposition $\vdash Q_C @ Q_B$, by exchanging its role with $Q_A$.

In conclusion, quantum teleportation looks like occurring inside a QC, without any need of an external space background, and then it is testable only by an internal observer. But this happens because the external space is isomorphic to the QC state space, and any external observer would appear as internal.
A scheme enlightening that difference between entanglement and quantum teleportation is given below.

**Entanglement** **Q-teleportation**

**Q-metalanguage** **Q-metalanguage**

EST is a disconnected space

$\vdash A, B$ and $\vdash A^{\perp}, B^{\perp}$ $\qquad \vdash (Q_A @ Q_B), Q_C$

↓ ↓

**Quantum logic** **Quantum logic**

The QC state space
is disconnected because of $Q_C$

(in the QC state space
isomorphic to the
fuzzy sphere )
The EST and the fuzzy sphere
**are not**

The EST and the QC state space
**are naturally** isomorphic:
no need of a logical reflection between
QML and QOL

Need of a logical reflection
Between QML and QOL

**Entanglement:** $\vdash Q_A @ Q_B$ **Teleportation:** $\vdash Q_C @ Q_B$

↓ ↓

The **ENT-theorem**
**is not** a meta-theorem

The **TEL-theorem**
**is** a meta-theorem

**4.5 The quantum simulation**

The difference between the topologies of the EST and the PC state space of an EPR pair is what does not allow to remain in the meta-logical domain, and requires an external observer.
But this in not enough. To describe the EST on the fuzzy sphere, which is isomorphic to the PC quantum state, it is also necessary to make a quantum simulation of EST through spin networks onto the fuzzy sphere. This is due to the fact that the representation of one qubit state in the EST, which is one Planck pixel, corresponds to two elementary cells in the fuzzy sphere. Spin networks as quantum simulators, introduced by Rasetti and Marzuoli [25], are a realization of quantum simulation in the true spirit of Feynman original idea [26]. The particular case of quantum simulation of the EST is under study [20]. We just give the conceptual scheme below.

**The Quantum Simulation**

**Fig.1**
**The EST**
Each pixel is punctured by a spin $j = \pm 1/2$

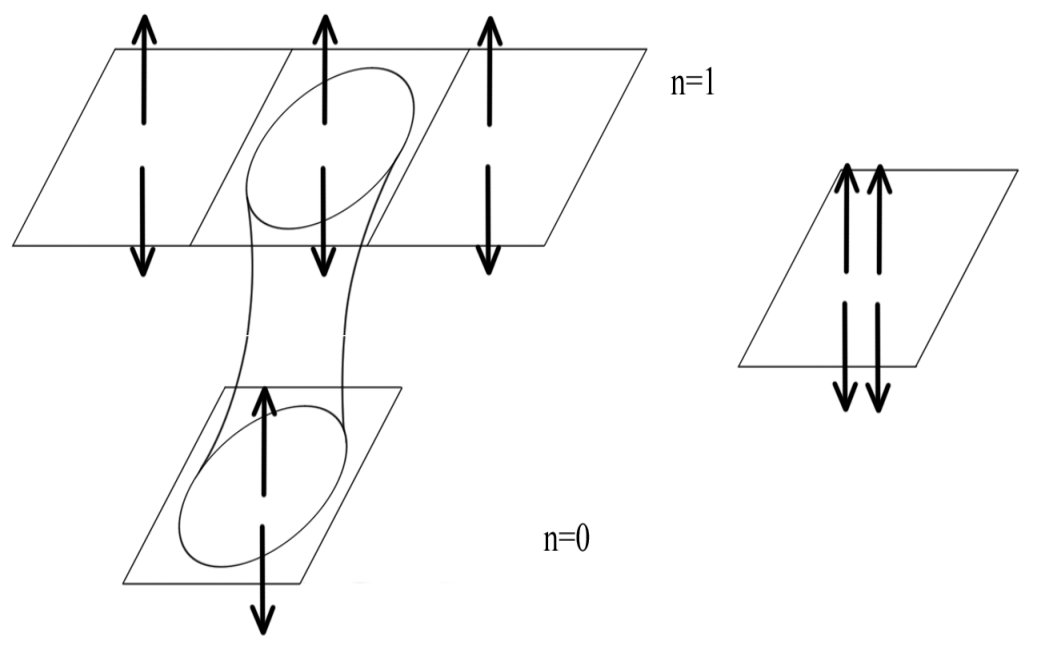

Without a quantum simulation, this would correspond to a fuzzy sphere with two cells, each cell encoding a bit. In this case, the fuzzy sphere describes one qubit state, thus cannot be used for describing entanglement.

**The fuzzy sphere in the minimal irreducible representation of SU(2)**

**Fig. 2**
A fuzzy sphere with $j = \pm 1/2$, n=2 cells

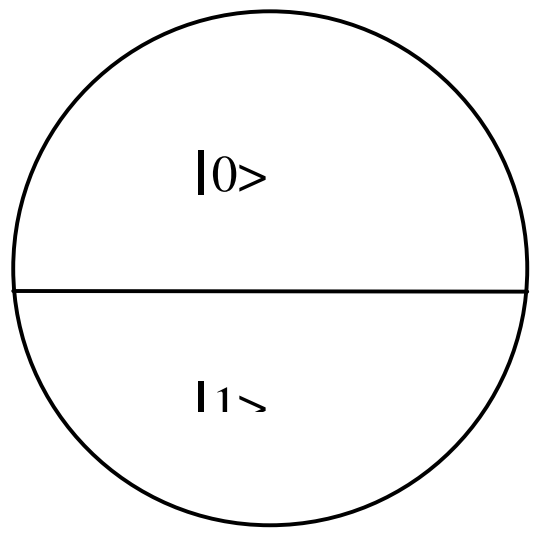

↓**Q-simulation of EST through spin networks**

Upon q-simulation, the EST will be described by a fuzzy sphere in the $j = 3/2$ irreducible representation of SU(2), but with 2 cells instead of 4. The two cells encode a Bell state. Note that 4 cells would correspond to 1 pixel at time $t_0$ plus 3 pixels at time $t_1$ of the original QGN without entanglement. Here the two cells of the fuzzy sphere correspond to the two pixels of the EST, one at time $t_0$, the other at time $t_1$, which are entangled through a mini wormhole.

**Fig. 3**
$j = \pm 3/2$, n=2 cells
The 2 cells encode a Bell state

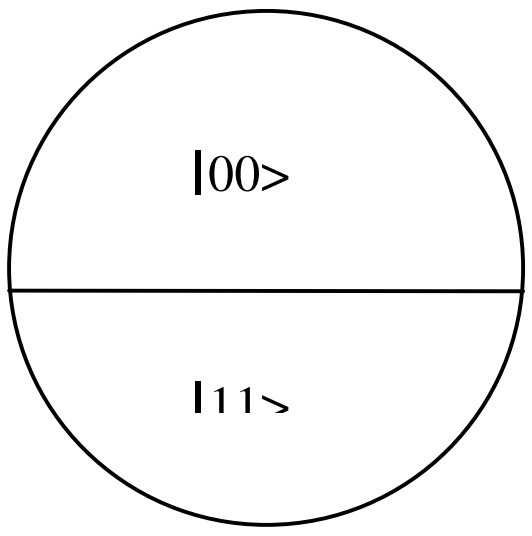

**↓isomorphism**

**QC space state**

The disconnected QC state space in the case of TEL is given in Fig. 4 and Fig. 5 in terms of two distinct fuzzy spheres. We remind that in this case there is no need of a logical reflection from QML to QOL.

**Fig. 4**
**Logical premises, initial state:**
**(Qubit $Q_A$ entangled with qubit $Q_B$)**
**times the unknown qubit state $Q_C$.**
**The quantum space is disconnected.**

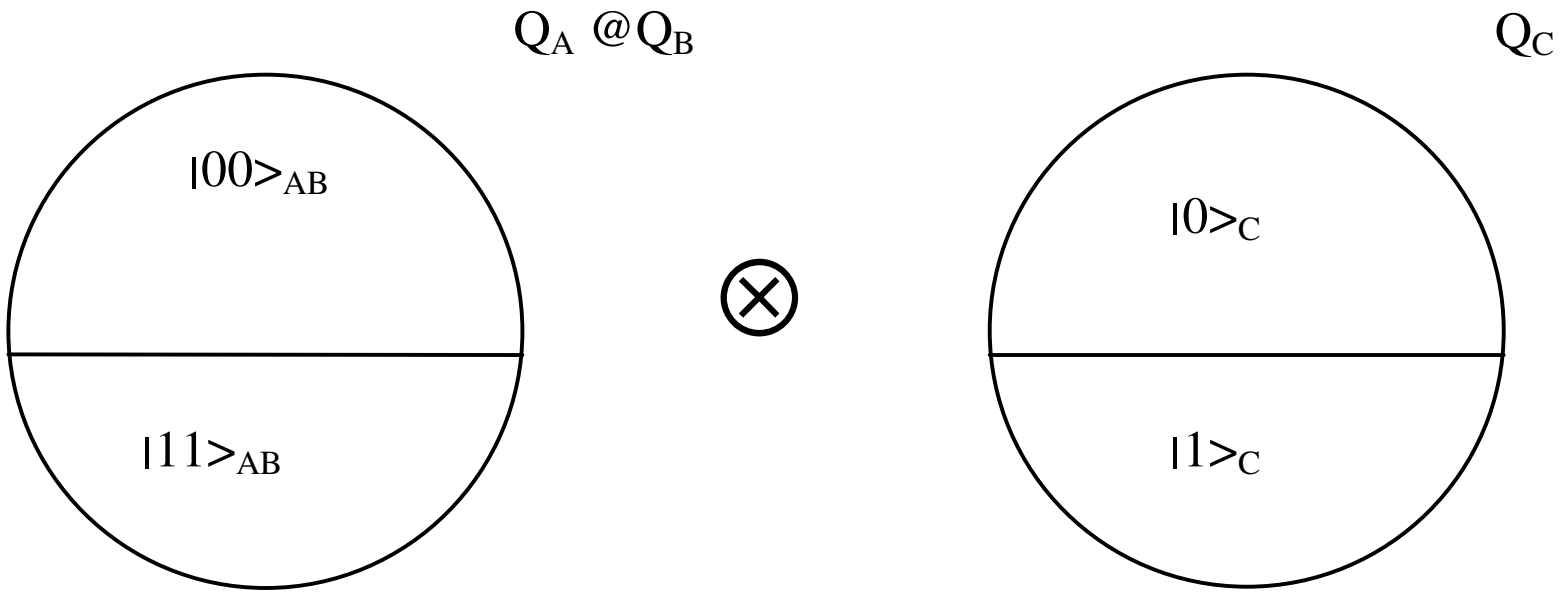

**↓Cut $Q_A$**

**Fig. 5**
**EPR 1-rule: qubit $Q_B$**
**with bit $|0>_C$ in parallel with**
**EPR 2-rule: qubit qubit $Q_B$**
**with bit |1>C. The result is entanglement**
**of qubit $Q_B$ with qubit $Q_C$.**

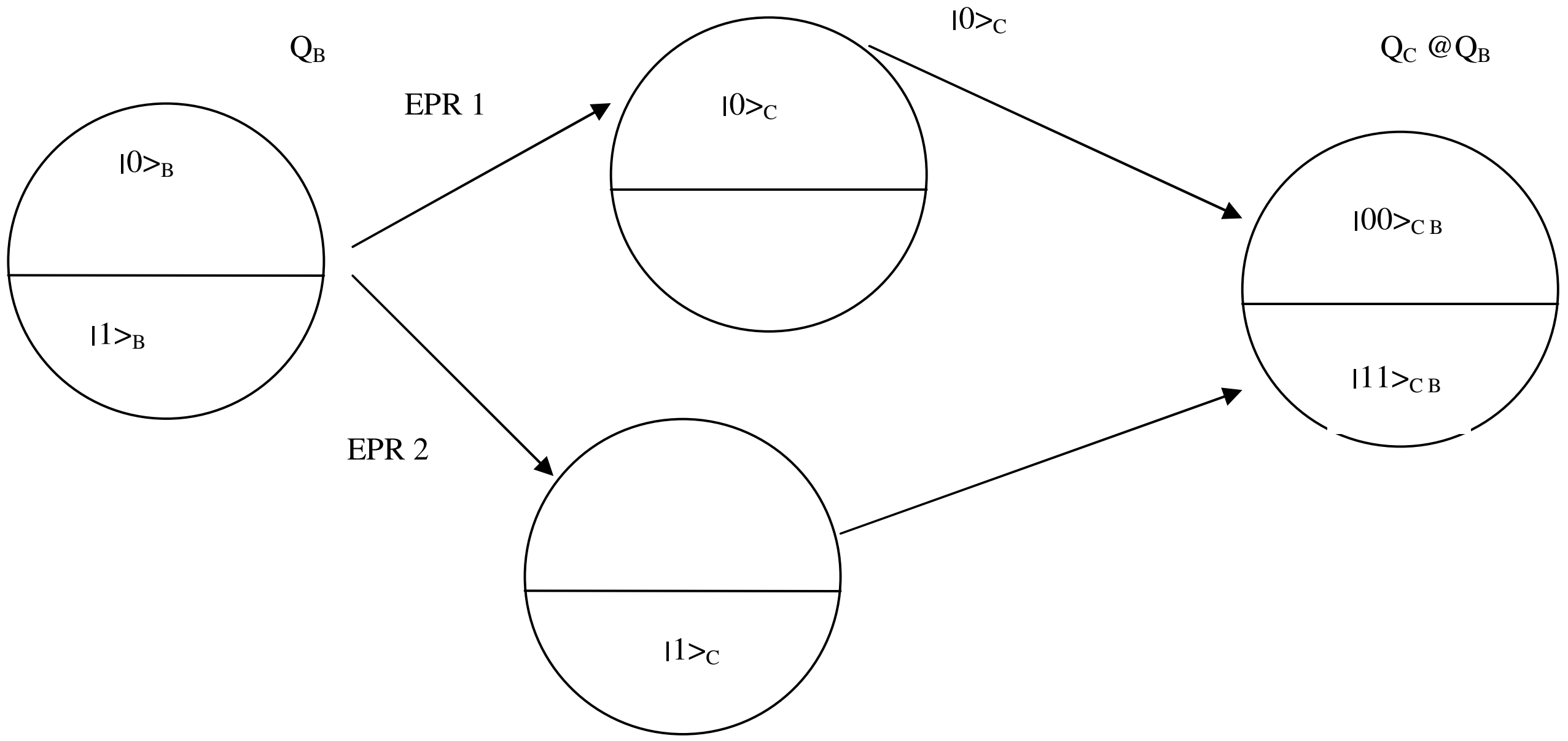

**4.6 To be, to become, or to meta-be**

Any quantum operation, reversible or irreversible, leads out of the meta theory. However, if this operation is performed in parallel, one remains in the meta theory, and there is not any substantial effect in the logical domain which should describe the quantum operation. Then, a quantum logical theorem when demonstrated in parallel gives back the premises in the conclusions. On the other hand, a quantum meta theorem can be demonstrated only with meta rules and with rules that reproduce quantum operations in parallel.

If we associate the metaphor of the internal/external observer with the fact of remaining or not in meta theory, the internal observer is nothing but performing a quantum operation (be it a projective measurement or a quantum and therefore reversible logical gate) in parallel. The result will always be the same: not having done anything. In fact, we remind that two projective measurements in parallel do not measure the qubit, because they do not destroy the quantum superposition, and two C-NOT quantum gates in parallel report to a separable state. This means that quantum entanglement, which is described by a quantum logical theorem, needs an external observer. On the contrary, quantum teleportation, which is logically described by a meta-theorem, allows only an internal observer.

The consequence in logic is that while the proof of the ENT-theorem can be given to a QC, the proof of the TEL-theorem cannot, because these theorems stand on two different logical levels. In Physics this leads to the conclusion that entanglement and quantum teleportation are two features of quantum computing which are ontologically quite different.

Ontology concerns questions regarding what entities exist or can come to existence. It needs not to be a material existence. For example, abstractions can be regarded as entities. A very important abstraction, in our case, is the passage (logical "reflection") from QML to QOL [2]. What is real in this case, the QML or the QOL? Or just the process of abstraction passing from the former to the latter? In an ontology of events, one would answer that what is real is the QOL. In an ontology of processes, the answer would be: the abstraction from QML to QOL. This is in fact the ontology of processes.

In an ontology of processes (to become) what is real in our case is the abstraction from QML to QOL. In an ontology of events (to be) what is real is the "event" in the QOL coming from such a process, and entanglement is an "event" . An event in the QOL can be observed by an external observer but not by an internal one (by internal we mean internal to the quantum system logically described by the QOL).

On the other hand, it is very difficult to define quantum teleportation either as a process or as an event. As a process, it does not come from the above abstraction, that is the reflection of QML into the QOL. As an event it does not belong to a QOL, and cannot be observed by an external observer.

In general, in the case of a quantum meta-theory, there is not a border between QML and QOL. Then, an ontology of processes does not apply in this case. Nothing can "become" to existence. However, it might be possible to consider an ontology of meta-events ("meta-be"), which are internal to the QML.

Then, quantum teleportation might be defined as a meta-event inside the QML. See the illustrative scheme below.

| **Ontology of processes** | **Ontology of events** | **Ontology of meta events** |
|---|---|---|
| (To become) | (To be) | (To meta-be) |
| **QML** | **QOL** | **QML** |
| ↓ | **Theorems** | **Meta theorems** |
| | Premises | Premises |
| **QOL** | 1) No branched proofs | Branched proofs only |
| | ↓ | ↓ |
| | Conclusions | Conclusions |
| | **External observer** | **Internal Observer** |
| | 2) Branched proofs | |
| | ↓ | |
| | Conclusions = premises | |
| | **Internal observer** | |

## 5. Conclusions

From the above results, entangled space-time seems to be endowed with the quantum meta-language necessary to control the quantum object-language of a quantum computer. Then a question arises: is it possible to perform quantum computation just by means of a quantum meta-language, without any material support to physically realize the quantum computer, which instead uses the quantum object-language? Or, more explicitly, can indeed the entangled space-time compute?
In the case of an empty quantum space-time, like the one we are considering, there is the possibility of self-programming, but in a very abstract way, that is, by using only the logical qubits encoded (by the quantum holographic principle [6]) in the quantum space-time itself. In other words, an empty and fully entangled quantum space-time is a kind of quantum software, which is able to program a possible quantum physical universe. In fact, if a particle should fill that space-time, it would get entangled with the latter, and consequently it will be described in terms of a quantum meta-language as well. As we have seen, this happens in the case of quantum teleportation of an unknown quantum state when conveyed by EST. The unknown quantum state of the particle is teleported from a site to another of the EST, and finally gets entangled with its destination site. The disembodied quantum information of the particle became part of the quantum space-time software.
In this paper, we used two meta-rules (the cut and the EPR rules) and one meta-theorem, the (TEL)-theorem. Note that the proof of the meta-theorem was built just by the use of the two meta-rules. This is a fully quantum meta-logical structure, which reveals the nature of entangled space-time as a quantum meta-language. In the case of quantum teleportation there is an identification between the quantum metalanguage of the EST and the quantum logic of the quantum computer (QC). This is due to a "natural" isomorphism between the EST and the QC state space in the case of quantum teleportation.. Although the proof of the (TEL)-theorem cannot be given to the QC, as it is a meta-theorem, quantum teleportation occurs as inside a quantum computer, without any reference to a physical quantum space-time background.
On the contrary, entanglement requires a logical "reflection" of the QML endowed by the EST into the QOL of the QC. The reflection allows the QC to use the program. But the two languages are not identified. For this reason the (ENT)-theorem is not a meta-theorem.
We conclude with two further remarks. The first one is that the quantum meta-language inherited by the entangled space-time supports the most salient features of quantum computing. The second one concerns the fact that the empty entangled space-time is programming its own geometrical structure by means of the endowed quantum meta-language. These two remarks together lead to the following conclusions:
the geometrical structure of quantum space-time is quantum computational.
Finally, we wish to stress the fact that the need of a "reflection" from QML to QOL is just due to the different topologies of the two quantum spaces by which the two languages are endowed.

**Acknowledgements**
I am very grateful to Mario Rasetti and Gianfranco Minati for useful comments and advice.

# Appendix

**The logical connective "Quantum superposition"**

The quantum logical connective ${}_{\alpha}\&_{\beta}$ was introduced [2] by the use of the reflection principle [10], which projects the metalanguage into the logical language. The definitional equation of ${}_{\alpha}\&_{\beta}$ is:

$$\Gamma|\!- C^{\perp}\,{}_{\alpha}\&_{\beta}\,C \quad \underline{\text{iff}} \quad \Gamma|\!-^{\alpha} C^{\perp} \quad \underline{\text{and}} \quad \Gamma|\!-^{\beta} C \tag{A.1}$$

where the upper fixes $\alpha$ and $\beta$ of the sequents in the RHS of Eq. (A.1) are the assertion degrees of the two atomic assertions.

The formation rule of the connective ${}_{\alpha}\&_{\beta}$ is:

$$\frac{\Gamma|\!-^{\alpha} C^{\perp} \qquad \Gamma|\!-^{\beta} C}{\Gamma|\!- C^{\perp}\,{}_{\alpha}\&_{\beta}\,C}\ \&-form. \tag{A.2}$$

**The cut rule in Basic logic**

In Basic logic [10], where it holds the principle of visibility, the cut rule holds in absence of active contexts:

$$\frac{\Gamma|\!- A \qquad A|\!- B}{\Gamma|\!- B}\ . \tag{A.3}$$

Let us consider the reflection rule for & in Basic logic:

$$\frac{B|\!-\Delta}{B\,\&\,A|\!-\Delta} \qquad \frac{A|\!-\Delta}{B\,\&\,A|\!-\Delta}(\&L) \tag{A.4}$$

and the same rule in intuitionistic logic [22] (which has full context on the left only) is:

$$\frac{\Gamma, A|\!-\Delta}{\Gamma, A\,\&\,B|\!-\Delta} \qquad \frac{\Gamma, B|\!-\Delta}{\Gamma, A\,\&\,B|\!-\Delta}(\&L)\ . \tag{A.5}$$

Now, if we replace $B$ with $A^{\perp}$ in both of them, we get respectively:

$$\frac{A^{\perp}|\!-\Delta}{A^{\perp}\,\&\,A|\!-\Delta} \qquad \frac{A|\!-\Delta}{A^{\perp}\,\&\,A|\!-\Delta}(\&L) \tag{A.6}$$

and:

$$\frac{\Gamma, A|\!-\Delta}{\Gamma, A\,\&\,A^{\perp}|\!-\Delta} \qquad \frac{\Gamma, A^{\perp}|\!-\Delta}{\Gamma, A\,\&\,A^{\perp}|\!-\Delta}(\&L)\ . \tag{A.7}$$

In Eq. (A.7) there is a chance that the superposed state $A\,\&\,A^{\perp}$ gets entangled with the environment $\Gamma$, instead in Eq. (A.6) there is not such a possibility. It is straightforward to show that the cut rule (2.2) in Basic logic would destroy the superposed state $|\!- A\,\&\,A^{\perp}$.

In fact, by replacing $A$ by $A\,\&\,A^{\perp}$, $B$ by $A$, and putting $\Gamma = \emptyset$ in Eq. (2.2), we get:

$$\frac{|\!- A\,\&\,A^{\perp} \qquad \dfrac{A|\!- A}{A\,\&\,A^{\perp}|\!- A}(\&L)}{|\!- A}(cut)\ . \tag{A.8}$$

By replacing $A$ by $A^{\perp}$, in Eq. (A.7) we would get, similarly:

$$\dfrac{\vdash A \,\&\, A^{\perp} \qquad \dfrac{A^{\perp} \vdash A^{\perp}}{A \,\&\, A^{\perp} \vdash A^{\perp}}(\&L)}{\vdash A^{\perp}}(cut) \quad . \qquad \text{(A.9)}$$

In physical terms Eq. (A.8) means that, given the cat state $|Q\rangle_A = \frac{1}{\sqrt{2}}\left(|0\rangle_A + |1\rangle_A\right)$ ($\vdash A \,\&\, A^{\perp}$), and performing the projective measurement on the bit $|1\rangle_A$ ($A \,\&\, A^{\perp} \vdash A$), we get as a result the bit $|1\rangle_A$ ($\vdash A$ in the conclusion). Similarly, Eq. (A.9) means that given the cat state, and performing the projective measurement on the bit $|0\rangle_A$ ( $A \,\&\, A^{\perp} \vdash A^{\perp}$), we get as a result the bit $|0\rangle_A$ ($\vdash A^{\perp}$ in the conclusion).

**The cut in parallel**

$$\dfrac{\dfrac{\vdash A\&A^{\perp} \qquad \dfrac{A \vdash A}{A\&A^{\perp} \vdash A}(\&L)}{\vdash A}(cut) \qquad \dfrac{\vdash A\&A^{\perp} \qquad \dfrac{A^{\perp} \vdash A^{\perp}}{A\&A^{\perp} \vdash A^{\perp}}(\&L)}{\vdash A^{\perp}}(cut)}{\vdash A\&A^{\perp}}(\&R) \qquad \text{(A.10)}$$

**The definitional equation of the connective @**

The quantum logical connective @ is introduced by solving the definitional equation:

$$\underline{\text{iff}} \qquad \Gamma \vdash A, B \quad \underline{\text{and}} \quad \Gamma \vdash A^{\perp}, B^{\perp} \quad . \qquad \text{(A.11)}$$

It should be noticed that, on the right hand side of the definitional equation, each of the two commas is reflected into a $\wp$ ($\wp$ reflects the <u>and</u> on the right, inside the sequent) while the meta-linguistic link <u>and</u> is reflected into & (& reflects <u>and</u> on the right, outside the sequent). Thus the connective @ is an additive as well as multiplicative conjunction which reflects two kinds of <u>and</u> on the right, one inside the sequent and the other outside the sequent.

Finally, the connective @ , is a derived connective (as it is expressed in terms of two other connectives, namely, $\wp$ and &), which, nevertheless, is defined by its own definitional equation.

By exchanging $A$ with $A^{\perp}$, the connective entanglement @ relates to the sequent $\vdash \left(A \wp B^{\perp}\right) \& \left(A^{\perp} \wp B\right)$, corresponding to the Bell state $|\Psi_{+}\rangle_{AB} = \frac{1}{\sqrt{2}}\left(|1\rangle_A |0\rangle_B + |0\rangle_A |1\rangle_B\right)$.

In this case the definitional equation of @ reads:

$$\vdash Q_A \,@\, Q_B \qquad \underline{\text{iff}} \qquad \vdash A^{\perp}, B \quad \underline{\text{and}} \quad \vdash A, B^{\perp} \qquad \text{(A.12)}$$

and the @ -formation rule is:

$$\text{@-formation} \quad \dfrac{\Gamma \vdash A^{\perp}, B \qquad \Gamma \vdash A, B^{\perp}}{\Gamma \vdash Q_A \,@\, Q_B} \quad . \qquad \text{(A.13)}$$

This version of the @-formation rule will come in handy for the "branched" proof of the entanglement theorem, which, as we will see, gives back a separable state in the conclusion.

**The @-rules and properties**

Solving the definitional equation for @ leads to the following rules:

@-formation $\dfrac{\Gamma \vdash A, B \qquad \Gamma \vdash A^{\perp}, B^{\perp}}{\Gamma \vdash Q_A \,@\, Q_B}$ . (A.14)

@-implicit reflection $\dfrac{\Gamma \vdash Q_A \,@\, Q_B}{\Gamma \vdash A, B} \qquad \dfrac{\Gamma \vdash Q_A \,@\, Q_B}{\Gamma \vdash A^{\perp}, B^{\perp}}$ . (A.15)

@-axioms $Q_A \,@\, Q_B \vdash A, B \qquad (Q_A \,@\, Q_B \vdash A^{\perp}, B^{\perp}$ . (A.16)

@-explicit reflection $\dfrac{A \vdash \Delta \qquad B \vdash \Delta'}{Q_A \,@\, Q_B \vdash \Delta, \Delta'} \qquad \dfrac{A^{\perp} \vdash \Delta \qquad B^{\perp} \vdash \Delta'}{Q_A \,@\, Q_B \vdash \Delta, \Delta'}$ . (A.17)

We give a list of the properties of the connective @. More details and proofs can be found in [2].

1) **Commutativity**: $Q_A \,@\, Q_B = Q_B \,@\, Q_A$

2) **Semi-distributivity**:

From the definitional equation of @ with $\Gamma = \varnothing$, that is:

$\vdash Q_A \,@\, Q_B$ iff $\vdash A, B$ and $\vdash A^{\perp}, B^{\perp}$

we get:

$(A \,\&\, A^{\perp}) \,@\, (B \,\&\, B^{\perp}) \equiv (A \wp B) \,\&\, (A^{\perp} \wp B^{\perp})$

by which we see that two terms are missing, namely $(A \wp B^{\perp})$ and $(A^{\perp} \wp B)$, so that @ has distributivity with absorption, which we call semi-distributivity.

3) **Associativity**

To discuss associativity of @, a third qubit $Q_C$ is needed, and $Q_A \,@\, (Q_B \,@\, Q_C) \equiv (Q_A \,@\, Q_B) \,@\, Q_C$ cannot be demonstrated in Basic logic, as $Q_C$ acts like a context on the right.

**The EPR meta-rule**

$$\dfrac{\dfrac{\dfrac{\Gamma \vdash Q_A \,@\, Q_B \qquad Q_A \vdash A}{\Gamma \vdash A \,@\, Q_B}}{\Gamma \vdash A, B}\,@-impl.refl.}{\Gamma \vdash A \wp B}\,\wp-form. \tag{A.18}$$

where semi-distributivity: $A \,@\, Q_B \doteq A, B$ has been used in the step "@-impl.refl".

**The EPR rule in parallel**

$$\dfrac{\dfrac{\dfrac{\vdash Q_A \,@\, Q_B \qquad Q_A \vdash A}{\vdash A \,@\, Q_B}\,cut}{\vdash A, B}\,@-impl.refl. \qquad \dfrac{\dfrac{\vdash Q_A \,@\, Q_B \qquad Q_A \vdash A^{\perp}}{\vdash A^{\perp} \,@\, Q_B}\,cut}{\vdash A^{\perp}, B^{\perp}}\,@-impl.refl}{\vdash Q_A \,@\, Q_B}\,@-form. \tag{A.19}$$

**Derivation of the H-rule**

$$H-rule$$

$$\dfrac{\dfrac{\vdash A^{\perp}}{\vdash A^{\perp}\;{}_{\frac{1}{\sqrt{2}}}\&_{\frac{1}{\sqrt{2}}}\;A}\,H-rule}{\vdash^{\frac{1}{\sqrt{2}}} A^{\perp} \qquad \vdash^{\frac{1}{\sqrt{2}}} A}\,\&-refl. \qquad \dfrac{\dfrac{\vdash A}{\vdash A^{\perp}\;{}_{\frac{1}{\sqrt{2}}}\&_{-\frac{1}{\sqrt{2}}}\;A}\,H-rule}{\vdash^{\frac{1}{\sqrt{2}}} A^{\perp} \qquad \vdash^{-\frac{1}{\sqrt{2}}} A}\,\&-refl. \quad . \tag{A.20}$$

$$H^{-1}-rule$$

$$\dfrac{\dfrac{\vdash^{\frac{1}{\sqrt{2}}} A^{\perp} \qquad \vdash^{-\frac{1}{\sqrt{2}}} A}{\vdash A^{\perp}\;{}_{\frac{1}{\sqrt{2}}}\&_{\frac{1}{\sqrt{2}}}\;A}\,\&-form.}{\vdash A^{\perp}}\,H^{-1}. \qquad \dfrac{\dfrac{\vdash^{\frac{1}{\sqrt{2}}} A^{\perp} \qquad \vdash^{-\frac{1}{\sqrt{2}}} A}{\vdash A^{\perp}\;{}_{\frac{1}{\sqrt{2}}}\&_{-\frac{1}{\sqrt{2}}}\;A}\,\&-form.}{\vdash A}\,H^{-1}. \quad . \tag{A.21}$$

$$H-rule\ in\ parallel$$

$$\dfrac{\dfrac{\dfrac{\vdash A^{\perp}}{\vdash A^{\perp}\;{}_{\frac{1}{\sqrt{2}}}\&_{\frac{1}{\sqrt{2}}}\;A}\,H-rule}{\vdash^{\frac{1}{\sqrt{2}}} A^{\perp} \qquad \vdash^{\frac{1}{\sqrt{\ }}} A}\,\&-refl. \qquad \dfrac{\dfrac{\vdash A}{\vdash A^{\perp}\;{}_{\frac{1}{\sqrt{2}}}\&_{-\frac{1}{\sqrt{2}}}\;A}\,H-rule}{\vdash^{\frac{1}{\sqrt{2}}} A^{\perp} \qquad \vdash^{-\frac{1}{\sqrt{2}}} A}\,\&-refl.}{\vdash A^{\perp}\,\&\,A^{\perp} \equiv\; \vdash A^{\perp}}\,\&-form. \quad \cdot \tag{A.22}$$

**Derivation of the C-NOT rule**

Let us consider the premise $B, A \vdash$ and apply to $A$ the negation ( $\neg$ )- formation rule $\dfrac{A\vdash}{\vdash A^{\perp}}\neg form$. We get:

$$B,\dfrac{\dfrac{B,A\vdash}{A\vdash}}{\dfrac{\vdash A^{\perp}}{\vdash B, A^{\perp}}}\neg form. \quad . \tag{A.23}$$

From the above sequent we get the first part of the CNOT rule:

$$B, A \vdash B, A^{\perp}\ . \tag{A.24}$$

The second part of the CNOT rule is obtained by performing the negation of both the antecedent, and of the consequent, and then proceeding by absurd, that is:

$$\neg - form\,\frac{B, A \vdash}{\vdash B^{\perp}, A^{\perp}} \tag{A.25}$$

$$\neg - refl\,\frac{\vdash B, A^{\perp}}{B^{\perp}, A \vdash} \tag{A.26}$$

Which is a paradox, because the negation of the premise of (A.23), that is (A.25), does not gives back the negation of the consequent, unless we replace $A$ with $A^{\perp}$ in the consequent of (A.26) .This replacement can be achieved by applying the $\neg$-form to A in the consequent of (A.26).

$$\dfrac{B^{\perp}, \dfrac{\dfrac{\vdash B, A^{\perp}}{A \vdash}}{\vdash A^{\perp}} \neg\ form.}{\vdash B^{\perp}, A^{\perp}} \tag{A.27}$$

which gives rise to the second part of the CNOT-rule:

$$B^{\perp}, A^{\perp} \vdash B^{\perp}, A^{\perp}\ . \tag{A.28}$$

**The C-NOT rule action**

$$\dfrac{\dfrac{\vdash Q_B, A^{\perp} \qquad Q_B \vdash B \qquad Q_B \vdash B^{\perp}}{\dfrac{\vdash B, A^{\perp}}{\vdash B, A} CNOT \qquad \dfrac{\vdash B^{\perp}, A^{\perp}}{\vdash B^{\perp}, A^{\perp}} CNOT}}{\vdash Q_B\ @\ Q_A} @- form \tag{A.29}$$

**The C-NOT rule in parallel**

$$\dfrac{\dfrac{\vdash Q_B, A^{\perp} \qquad Q_B \vdash B \qquad Q_B \vdash B^{\perp} \qquad \vdash Q_B, A}{\dfrac{\vdash B, A^{\perp}}{\vdash B, A} CNOT \quad \dfrac{\vdash B^{\perp}, A^{\perp}}{\vdash B^{\perp}, A^{\perp}} CNOT \quad \dfrac{\vdash B, A}{\vdash B, A^{\perp}} CNOT \quad \dfrac{\vdash B^{\perp}, A}{\vdash B^{\perp}, A} CNOT}}{\vdash Q_B, Q_A} \&-form \tag{A.30}$$

**Proof of the (ENT)-theorem**

$$\dfrac{\dfrac{\dfrac{\dfrac{\vdash Q_B, Q_A \qquad Q_A \vdash A^{\perp}}{\vdash Q_B, A^{\perp} \qquad Q_B \vdash B \quad Q_B \vdash B^{\perp}} cut}{\vdash B, A^{\perp} \qquad \vdash B^{\perp}, A^{\perp}} \& -refl.axioms}{\vdash B, A \qquad \vdash B^{\perp}, A^{\perp}} CNOT}{\vdash Q_B\ @\ Q_A} @- form \tag{A.31}$$

We remind that the premises in the first sequent are the separable state of two qubits $\vdash Q_B, Q_A$ , and the projective measurement of qubit $Q_A$ to the zero bit $A^{\perp}$, that is $Q_A \vdash A^{\perp}$ . The cut rule then "cuts" the qubit $Q_A$ , and in the second sequent we have the separable state of the qubit $Q_B$ and the bit $A^{\perp}$, namely $|\!-Q_B, A^{\perp}$, plus the &-reflection axioms $Q_B|\!-B \quad Q_B|\!-B^{\perp}$. Then, by applying the distributive property of the multiplicative conjunction $\wp$ with respect to &, we are left with the two-bits states $|\!-B, A^{\perp}$ and $|\!-B^{\perp}, A^{\perp}$ in the third sequent. Then, the CNOT rule is applied, and in the fourth sequent, we get the premises for the formation rule of the connective entanglement @ . The conclusion is the entangled state $|\!-Q_B \;@\; Q$.

**Proof of the no-go theorem: "No-Entanglement in parallel"**

$$\dfrac{\dfrac{\dfrac{\dfrac{|\!-Q_B,Q_A \quad Q_A|\!-A^{\perp}}{|\!-Q_B,A^{\perp} \quad Q_B|\!-B \quad Q_B|\!-B^{\perp}}\,cut}{|\!-B,A^{\perp} \quad |\!-B^{\perp},A^{\perp}}\,\&\!-\!refl}{-B,A \quad |\!-B^{\perp},A^{\perp}}\,CNOT \qquad \dfrac{\dfrac{\dfrac{|\!-Q_B,Q_A \quad Q_A|\!-A}{|\!-Q_B,A \quad Q_B|\!-B \quad Q_B|\!-B^{\perp}}\,cut}{|\!-B,A \quad |\!-B^{\perp},A}\,\&\!-\!refl}{-B,A^{\perp} \quad |\!-B^{\perp},A}\,CNOT}{\dfrac{\dfrac{|\!-B,(A\&A^{\perp}) \quad |\!-B^{\perp},(A\&A^{\perp})}{|\!-(B\&B^{\perp}),(A\&A^{\perp})}\,\&\!-\!forn}{|\!-Q_B,Q_A}}\,\&\!-\!form \quad . \tag{A.32}$$

Note that in the left-side branch, the target in the CNOT is $|0\rangle_A$, corresponding to $A^{\perp}$, while in the right-side branch is $|1\rangle_A$, corresponding to $A$. Then, the conclusion in the left-side branch is the Bell state $|\Phi_+\rangle = \frac{1}{\sqrt{2}}(|00\rangle + |11\rangle)$, while the conclusion in the right-side branch is the Bell state $|\Psi_+\rangle = \frac{1}{\sqrt{2}}(|01\rangle + |10\rangle)$. The total conclusion is the sum of the two partial conclusions, which gives the tensor product $|Q\rangle_A \otimes |Q\rangle_B$, that is, a separable state, corresponding to the sequent $|\!-Q_A, Q_B$.

**Proof of the (TEL)-meta-theorem**

The proof of the (TEL)-meta-theorem is given below.

$$\dfrac{\dfrac{\dfrac{|\!-(\mathrm{Q_A} \;@\; \mathrm{Q_B}),Q_C \quad \mathrm{Q_A},Q_C|\!-^{\beta}\mathrm{C}}{|\!-^{\beta}C \;@\; Q_B}\,cut}{|\!-^{\beta}C,B}\,@\!-impl.refl. \qquad \dfrac{\dfrac{|\!-(\mathrm{Q_A} \;@\; \mathrm{Q_B}),Q_C \quad \mathrm{Q_A},Q_C|\!-{}^{\alpha}\mathrm{C}^{\perp}}{|\!-^{\alpha}C^{\perp} \;@\; Q_B}\,cut}{|\!-^{\alpha}C^{\perp},B^{\perp}}\,@\!-impl.refl}{|\!-Q_C \;@\; Q_B}\,@\!-form. \tag{A.33}$$

The appearance of the upper fixes $\alpha$ and $\beta$ in the above equation will be explained in what follows.

The unknown qubit $\mathrm{Q_C}$ hold by Alice at the beginning is the general qubit state $|\mathrm{Q}\rangle_C = \alpha|0\rangle + \beta|1\rangle$, where $\alpha$ and $\beta$ are complex numbers, which satisfy the relation: $|\alpha|^2 + |\beta|^2 = 1$. In the logic of qubits [2] it was necessary to introduce the connective "quantum superposition" $= {}_{\alpha}\&_{\beta}$ (which is the quantum analogous of the classical "and"=&) to take into account the probability amplitudes $\alpha$ and $\beta$. The logical qubit $\mathrm{Q_C}$ is then the compound proposition $\mathrm{Q_C} = C^{\perp}{}_{\alpha}\&_{\beta}C$. The atomic propositions $C^{\perp}$ and $C$ have partial truth values $|\alpha|^2$ and $|\beta|^2$ respectively, and the compound proposition $C^{\perp}{}_{\alpha}\&_{\beta}C$ has truth value 1.